\documentclass{article}
\usepackage{srcltx}
\usepackage[T2A]{fontenc}
\usepackage[utf8]{inputenc}
\usepackage[english]{babel}

\usepackage[a4paper,top=4cm,bottom=3.5cm,left=2.5cm,right=2.5cm,marginparwidth=1.75cm]{geometry}
\usepackage{graphicx}

\begin{document}

\title{{About the observational check of the mechanism of gamma radiation in  Soft Gamma Repeaters (SGR)}}

 \medskip

    \author{ G.S. Bisnovatyi-Kogan \footnote{{Space research institute RAS, Moscow.}}}

\maketitle

                 {\centerline{\bf{Abstract.}}}
\bigskip

 Soft gamma repeaters (SGR) are identified as single neutron stars (NS) inside the Galaxy, or nearby galaxies, with sporadic transient gamma radiation. A total number of discovered SGR, including relative Anomalous X-ray pulsars (AXP), is few tens of objects. Many of them show periodic radiation, connected with NS rotation, with periods 2-12 s. The slow rotation is accompanied by small rate of loss of rotational energy, which is considerably smaller than the observed sporadic gamma ray luminosity, and is many orders less that the luminosity during giant bursts, observed in 4 SGR. Therefore the energy source is usually connected with annihilation of very strong NS magnetic field. Another model is based on release of a nuclear energy stored in the NS nonequilibrium layer. We suggest here an observational test with could distinguish between these two models.


 \medskip

\section{Introduction}

 Soft gamma repeaters (SGR),are discovered observationally in the year 1979 \cite{mg79b,mg79c,gm79} as a giant burst with recurrent gamma flared, and soon were connected with the discovered short gamma ray bursts (GRB) \cite{mg82} as a separate class of GRB.
 They  are identified now as single neutron stars (NS) inside the Galaxy, or nearby galaxies, with sporadic transient gamma radiation. A total number of discovered SGR, including relative Anomalous X-ray pulsars (AXP), is few tens of objects \cite{ruf17}. Some of them show periodic radiation, connected with NS rotation, with periods 2-12 s.
It seems now that short GRB include two different types of objects: merging of binary NS at cosmological distances, and another class represented by giant bursts in SGR at neighbouring galaxies at distances, when  recurrent sporadic SGR radiation is too faint for registration at the Earth.
Such events had been discussed in \cite{bk99}, and were discovered
in the galaxies M31 (Andromeda) \cite{ma08}, and a galaxy in the M81 group of galaxies \cite{fp07b}, see also \cite{plp05,hbs05}.

 The slow rotation is accompanied by small rate of loss of rotational energy, which is considerably smaller than the observed sporadic gamma ray luminosity, and is many orders less that the luminosity during giant bursts, observed in 4 SGR
 \cite{mg79c,mc99c,mc05,ma99a}. Therefore the energy source is usually connected with annihilation of very strong NS magnetic field \cite{dt92,dt95}. Another model is based on release of a nuclear energy stored in the NS nonequilibrium layer \cite{bkch74,bkinch75,bk92,bk12}. We suggest here an observational test with could distinguish between these two models.

\section{Periodical pulses}

First discovery of FXP 0520-66 by Konus experiment \cite{mg79b,mg79c} was accompanied by observation of periodic pulsation in this source with the period $P\approx 8$ s. in the afterglow radiation of the giant burst, lasting about 200 s. Similar pulsations had been observed in the afterglows of giant bursts it the objects $SGR 1900+14$ with the period $P\approx 5.16$ s. \cite{mc99c,ks99,mg79a}, and SGR 1806-20 with the period $P\approx 7.47$ s. \cite{mc05,pb05}. The periodicity  in the radiation of this source was found earlier \cite{kd98} in one of the recurrent outbursts, indications to such periodicity had been found in \cite{uf93}, and first discovery of this source in the year 1986 was reported in \cite{lf86a,lf86b}. The source SGR 1806-20 was probably the first SGR, in which periodic pulsations had been discovered in one of the recurrent outbursts, which period was later confirmed in the afterglow of the giant burst 27 December 2004, which was the most powerful from four observed giant bursts. Periodic pulsations had been found in the recurrent outbursts of several other SGR  \cite{ruf17,re10,aec11,gbk20}.

Connection of SGR with highly magnetized slowly rotating NS (magnetars) meets several problems. They are connected with observation of low magnetic field SGR \cite{re10,aec11}. Observations of several, quite normal radio pulsars without traces of recurrent gamma outbursts, having magnetar-like periods and magnetic fields \cite{msk03} are not consistent with this model.
These problems are discussed in details in \cite{bk16,bk17}. It is evident, that magnetic origin of gamma ray outbursts in SGR implies a  different behaviour of pulsations, in comparison with nuclear explosion origin \cite{bkch81,bk12,bki14}.We suggest that this difference should be visible in the behaviour of  pulsation phases, what could be derived from observations.

\section{Phase analysis}

Phase analysis of pulsations in rotating magnetized NS, observed as pulsars (radio, optical, X-ray, gamma-ray) is a very powerful tool, permitting to clarify the properties of the sources. In the binary millisecond pulsars with extremely  precise time stability of pulses the timing analysis of pulses gave the best precision results in favour of confirmation of General Relativity (GR) as a true theory of gravitation (see e.g. \cite{bk06}). This stability is connected with a very stable NS rotation, with practically fixed encoring of the magnetic field to the NS body. Such properties permit to exclude the changes connecting with a motion of magnetic field over NS surface, and interpret the visible period variations in timing data to the pulsar motion in the binary system, and different GR effects.

   In SGR the precision of observation is incomparably worse than in radiopulsars, so only sufficiently strong phase shifting in observations could be noticed and measured.
   NS in SGR is considered as single object, so no  motion in binary, and very small GR effects are not be visible in their timing data. The reason for strong phase shift in the SGR pulsations could be expected due to motion of the the burst location over the NS surface from one burst to another one. Such motion is unlikely to be possible if the transient bursts are connected with magnetic events. Extremely high conductivity of the pulsar matter prevents any magnetic field  changes in static situation, and does not permit changes in the large scale magnetic field configuration formed by electrical currents situated deep inside, it quiet region of the star.
The large scale of the magnetic field should determine the position of the bursting spot in all cases, not permitting for its motion over the star surface.
   The shift of the explosion spot over the surface of the NS is possible in the case when the explosion  mechanism  is not connected directly to the magnetic field large scale structure. Such mechanism of explosion, connected with processes in the NS non-equilibrium layer \cite{bkch74} was considered in \cite{bk12,bki14,bk16,bk17}. Earlier it was suggested as a mechanism for formation of gamma ray bursts (GRB) inside the Galaxy \cite{bkinch75,bkch79}. Discovery of cosmological origin of GRB exclude this mechanism for GRB model, but it remains for application to SRG bursts. A considerable phase shift in the gamma ray pulsar timing, between the observed transient bursts, would be a strong argument in favour of the nuclear explosion SGR burst mechanism.

   \section{Phase shift observations in SGR}

   The pulsed radiation of SGR was first discovered in the famous 5 March 1979 burst \cite{mg79b,mg79c}. Subsequently it was visible very distinctly in the afterglows of two other giant bursts  SGR 1900+14 and SGR 1900+14. Analysis of many recurrent bursts in the last two objects had shown the periodic component in their radiation, with almost the same, slightly growing period.

     \subsection{SGR 1900+14}

The pulsations with a period 5.16 s. have been observed in the afterglow of the giant burst during several hundred of seconds.    \cite{mc99c,ks99}. Search for periodic pulsations had been done in recurrent bursts and persistent radiation of this object in subsequent years in the bands from radio \cite{lx00} to optics and hard X-rays \cite{wk99,xmm06,gmt06,emt07,irm08,tck12,tb19}. The pulsation with a corresponding  period have been observed in all bands, except radio.

   \subsection{SGR 1806-20}

 The pulsed signals with the period 7.47 s. had been discovered  in the year 1998 \cite{uf93,kd98}, 6 years before the giant burst in this SGR. After observations of the giant burst \cite{mc05,pb05}, periodic pulsations in this SGR had been visible in its radiation during subsequent observations in several X-ray bands \cite{ykk15,ybk17}.

   \subsection{SGR J1935+2154}

 SGR J1935+2154 was discovered in 2014 \cite{smp14}, and has a spin period $P \sim 3.25$  s.
 It is one of the most active magnetars, showing
outbursts almost every year \cite{ykj17,lgr20}. Its
reactivation date in 2020 April 27,  was observed by several X-ray and gamma-ray instruments when a burst storm and an increase of the persistent X-ray flux were detected \cite{ygk20,gbko20}.
Subsequent activity of this GRB during 7 months was followed by SMM and Chandra missions \cite{bzi22}.

\section{Conclusions}

The three above considered objects were observed during a long time, when several strong outbursts happen. They looks out as best candidate for making phase analysis, which should show the behaviour of the peak activity spot on the NS surface. If it does not change its position on the NS during all the time, than it is most probably connected with the large scale structure of the magnetic field, and witnesses in favour of the magnetar model. If its position changes after strong outburst that its activity is presumably determined by nuclear explosions, connected with existence of non equilibrium layer in  NS  envelope.

\section*{Acknowledgements}

The author is grateful A.S Pozanenko for discussions, and  N.R. Ikhsanov for useful comments.
This work is partially supported by RFBR grant 20-02-00455.

\end{document}